\documentclass[epj, nopacs]{svjour}
\usepackage{graphics}
\usepackage{amsmath}
\usepackage{listings}
\def\slash#1{
\setbox0=\hbox{$#1$}                    
\dimen0=\wd0                            
\setbox1=\hbox{/} \dimen1=\wd1          
\ifdim\dimen0>\dimen1                   
\rlap{\hbox to \dimen0{\hfil/\hfil}}    
#1                                      
\else                                   
\rlap{\hbox to \dimen1{\hfil$#1$\hfil}} 
/                                       
\fi}
\begin{document}
\title{Calculation of HELAS amplitudes for QCD processes using graphics
processing unit (GPU)}
\author{K.~Hagiwara\inst{1}
   \and J.~Kanzaki\inst{2}\fnmsep\thanks{e-mail: junichi.kanzaki@kek.jp}
   \and N.~Okamura\inst{2}\fnmsep\thanks{e-mail: naotoshi@post.kek.jp}
   \and D.~Rainwater\inst{3}
   \and T.~Stelzer\inst{4}\fnmsep\thanks{e-mail: tstelzer@uiuc.edu}
   }                     
   %
   %
\institute{KEK Theory Center and Sokendai, Tsukuba 305-0801, Japan
	\and KEK, Tsukuba 305-0801, Japan
	\and Space and Geophysics Laboratory, Applied Research
	Laboratories, University of Texas, Austin, TX 78758, USA
	\and Dept. of Physics, University of Illinois, Urbana, IL, USA
}

\date{Received: date / Revised version: \today}

\abstract{
We use a graphics processing unit (GPU) for fast calculations of helicity
amplitudes of quark and gluon scattering processes in massless QCD.
New HEGET  ({\bf H}ELAS {\bf E}valuation with {\bf G}PU {\bf E}nhanced
{\bf T}echnology) codes for gluon self-interactions are introduced, and
a C++ program to convert the MadGraph generated FORTRAN codes into HEGET
codes in CUDA (a C-platform for general purpose computing on GPU) is
created. 
Because of the proliferation of the number of Feynman diagrams and the
number of independent color amplitudes, the maximum number of final
state jets we can evaluate on a GPU is limited to 4 for pure gluon
processes ($gg\!\rightarrow\! 
4g$), or 5 for processes with one or more quark lines such as
$q\overline{q}\!\rightarrow\! 5g$ and $qq\!\rightarrow\! qq\!+\!3g$.
Compared with the usual CPU-based programs,  we obtain 60-100
times better performance on the GPU, except for 5-jet production
processes and the $gg\!\rightarrow\! 4g$ processes for which the GPU
gain over the CPU is about 20. 
} 
		
\titlerunning{Calculation of HELAS amplitudes for QCD}
%

\maketitle

 \section{Introduction}
 \label{sec:intro}
 
 In our previous report~\cite{qed-paper} we introduced a
 C-language~\cite{cuda} version of the HELAS codes~\cite{helas}, HEGET
 ({\bf H}ELAS {\bf E}valua\-tion with {\bf G}PU {\bf E}nhanced {\bf
 T}echnology), which can be used to compute helicity amplitudes on a GPU
 (Graphics Processing Unit).
 Encouraging results with 40-150 times faster computation speed over the
 CPU performance were obtained for pure QED processes, $q\overline{q}
 \!\rightarrow\! n\gamma$, for $n\!=\!2$ to 8 in $pp$ collisions.

 In this paper, we extend our study to QCD processes with massless
 quarks and gluons.
 The HEGET routines for massless quarks and gluons are identical to those
 of quarks and photons introduced in \cite{qed-paper}, and the $qqg$
 vertex function structure is also the same as the $qq\gamma$ functions.
 The only new additional routines are those for $ggg$ and $gggg$ vertices.
 For the QED processes studied in ref.~\cite{qed-paper}, we found that
 the present CUDA compiler cannot process $q\overline{q}\!\rightarrow\!
 6\gamma$ amplitude with $6!\!\approx\! 700$ Feynman diagrams, and we need
 to subdivide the HEGET codes into small pieces for $6\gamma$ and
 $7\gamma$ processes. 
 In the case of $8\gamma$ production with $8!\!\approx\! 4\times 10^{4}$
 Feynman diagrams, we have not been able to compile the program even
 after subdivision into small pieces.
 We also encountered serious slow down when the program accesses global 
 memory during the parallel processing period.
 Therefore, our concern for evaluating the QCD processes on a GPU is the
 proliferation of the number of diagrams,  as well as the  number of
 independent color amplitudes which come with different color weights.

 The paper is organized as follows.
 In section 2, we present the cross section formula for $n$-jet
 production processes in $pp$ collisions in the quark-parton model,  or
 in the leading order of perturbative QCD with scale-dependent parton
 distribution functions (PDF's).
 In section 3, we review briefly the structure of GPU computing by using
 HEGET codes, and give basic parameters of the GPU and CPU machines used
 in this analysis. 
 In section 4, we introduce new HEGET functions for $ggg$ and $gggg$
 vertices. 
 Section 5 gives our results and section 6 summarizes our findings.
 Appendix lists all the new HEGET codes introduced in section 4.

 \section{Physics Process}
 \label{sec:physics}

  \subsection{$n\textrm{-jet}$ production in $pp$ collisions}
   \begin{table*}[bt]
    \begin{center}
     \caption{The number of Feynman diagrams and the color bases for QCD
     processes studied in this paper.}
     \label{tab:qcd-processes}       
     \begin{tabular}{|c|c|c|c|c|c|c|} \hline
      No. of jets
      & \multicolumn{2}{|c|}{$gg\!\rightarrow\!\mathrm{gluons}$} 
      & \multicolumn{2}{|c|}{$u\overline{u}\!\rightarrow\!\mathrm{gluons}$} 
      & \multicolumn{2}{|c|}{$uu\rightarrow uu\!+\!\mathrm{gluons}$} \\
      \cline{2-7}
      in the final state
      & \#diagrams & \#colors & \#diagrams & \#colors & \#diagrams & \#colors \\
      \hline
      2 & 6 & 6 & 3 & 2 & 2 & 2 \\ \hline
      3 & 45 & 24 & 18 & 6 & 10 & 8 \\ \hline
      4 & 510 & 120 & 159 & 24 & 76 & 40 \\ \hline
      5 & 7245 & 720 & 1890 & 120 & 786 & 240 \\ \hline
     \end{tabular}
    \end{center}
   \end{table*}
  
  The cross section for $n$-jet production processes can be
  expressed as  
  \begin{eqnarray}
   {\rm d\sigma}=
    \sum_{\{a,b\}}\!\iint \mathrm{d}x_a \mathrm{d}x_b
    D_{a\!/\!p}\left(x_{a},\!Q\right)
    D_{b\!/\!p}\left(x_{b},\!Q\right)
    {\rm d\hat{\sigma}}(\hat{s})\,,
    \label{eq:process}
  \end{eqnarray}
  where $D_{a\!/\!p}$ and $D_{b\!/\!p}$ are the scale ($Q$) dependent
  parton distribution functions (PDF's),  $x_{a}$ and $x_{b}$ are the
  momentum fractions of the partons $a$ and $b$, respectively, in the right-
  and left-moving protons.
  For the total $pp$ collision energy of $\sqrt{s}$,
  \begin{eqnarray} 
   \hat{s}=s\, x_{a}\, x_{b}\, ,
  \end{eqnarray}
  gives the invariant mass squared of the hard collision process
  \begin{equation}
  \mathbf{a}+\mathbf{b}\rightarrow
   \mathbf{1}+ \mathbf{2}+ \cdots + \mathbf{n}\,.
  \end{equation}
  The subprocess cross section is computed in the leading order as
   \begin{eqnarray}
    {\rm d\hat{\sigma}}(\hat{s})
     &=&\dfrac{1}{2\hat{s}}\frac{1}{2\!\cdot\! 2}
     \sum_{\lambda_{i}} 
     \dfrac{1}{n_{a}n_{b}} 
     \sum_{c_{i}}
      \left|
       {\mathcal{M}}_{\lambda_{i}}
      ^{c_{i}}\right|^{2}
     {\rm d\Phi_{n}}\,,
   \end{eqnarray}
   where
   \begin{eqnarray}
    {\rm d\Phi_{n}}
     &=&
     \left(2\pi\right)^4
     \delta^4\left(\!p_{a}\!+\!p_{b}\!-\!\sum_{i=1}^{n}p_i\!\right)
     \prod_{i=1}^{n}
     \dfrac{\mathrm{d}^3 p_i}{(2\pi)^3\,2\omega_i}\,,
     \label{eq:phase-space}
   \end{eqnarray}
   is the invariant $n$-body phase space, $\lambda_{i}$
   are the  helicities of the initial and  final partons,
   $n_{a}$ and $n_{b}$ are the color degree of freedom of the initial
   partons, $a$ and $b$, respectively, and
   $c_{i}$ represents the color indices of the initial and
   final partons.
   When there are more than one gluons or identical quarks in the final
   states, an appropriate statistical factor should be multiplied on the
   phase space  $\mathrm{d\Phi_{n}}$ in eq.~(\ref{eq:phase-space}).

  The Helicity amplitudes for the process (\ref{eq:process})
  \begin{equation}
   {\renewcommand\arraystretch{1.5}
   \begin{array}{rcl}
    \lefteqn{\mathbf{a}(p_{a},\lambda_{a},c_{a}) +
    \mathbf{b}(p_{b},\lambda_{b},c_{b})} \\
    & \rightarrow &
    \mathbf{1}(p_{1},\lambda_{1},c_{1}) + \cdots
    + \mathbf{n} (p_{n},\lambda_{n},c_{n})
   \end{array} 
    }
  \end{equation}
  can be expressed as
  \begin{equation}
   \mathcal{M}_{\lambda_{i}}^{c_{i}}
    = \sum_{l\in{\rm diagram}} \left( M_{\lambda_{i}}\right)_{l}^{c_{i}}
    \label{eq:matrix-element}
  \end{equation}
  where the summation is over all the Feynman diagrams.
  The subscripts $\lambda_{i}$ stand for a given combination of
  helicities ($\pm 1$ for both quarks and gluons in the HELAS
  convention~\cite{helas}), and the subscripts $c_{i}$ correspond to a
  set of color indices (1, 2, 3 for flowing-IN quarks, $\overline{1},
  \overline{2}, \overline{3}$ for flowing-OUT quarks, and 1 to 8 for
  gluons). 
  In MadGraph~\cite{madgraph} the amplitudes are expanded as
  \begin{equation}
   \mathcal{M}_{\lambda_{i}}^{c_{i}}
    = \sum_{\alpha} T_{\alpha}^{c_{i}} (J_{\lambda_{i}})_{\alpha}
    \label{eq:color-amplitude}
  \end{equation}
  in the color bases $T_{\alpha}^{c_{i}}$ which are made from the SU(3)
  generators in the fundamental representation~\cite{su3}

  The color factors are computed as
  \begin{equation}
   \mathcal{N}_{\alpha\beta} = \frac{1}{n_{a}n_{b}} \sum_{c_{i}}
    (T_{\alpha}^{c_{i}}) (T_{\beta}^{c_{i}})^{*}
    \label{eq:color-matrix}
  \end{equation}
  where $n_{a,b}\! =\! 3$ for $q$ and $\overline{q}$, $n_{a,b}\!=\!8$
  for gluons, and the summation is over all $\{c_{i\}} =
  \{c_{a},c_{b},c_{1},\ldots,c_{n}\}$.
  The color sum-averaged square amplitudes are computed as
  \begin{equation}
   \overline{\sum_{c_{i}}}\,
    \left|\mathcal{M}_{\lambda_{i}}^{c_{i}}\right|^{2}
    = \sum_{a,b} (J_{\lambda_{i}})_{\alpha} {\mathcal{N}}_{\alpha\beta}
     (J_{\lambda_{i}})_{\beta}^{*}.
    \label{eq:color-averaged-amplitude}
  \end{equation}
  The cross sections are then expressed as
  \begin{equation}
   \mathrm{d}\hat{\sigma}(\hat{s}) = \frac{1}{2\hat{s}} \,
    \overline{\sum_{\lambda_{i}}}\;
    \overline{\sum_{c_{i}}}\,
    \left|\mathcal{M}_{\lambda_{i}}^{c_{i}}\right|^{2}\,
    \mathrm{d\Phi_{n}}\,,
  \end{equation}
  where we introduce the helicity sum-average symbol as
  \begin{equation}
   \overline{\sum_{\lambda_{i}}} \,\equiv\, \frac{1}{2}\frac{1}{2}
    \sum_{\lambda_{i}}\,.
  \end{equation}

  In this paper the following three types of multi-jet production
  processes are computed:
  \begin{subequations}
   \begin{eqnarray} 
    gg &\rightarrow& gg, ggg, gggg \\
    q\overline{q} &\rightarrow& gg, ggg, gggg, ggggg \\
    uu &\rightarrow& uu, uug, uugg, uuggg
   \end{eqnarray}
   \label{eq:processes}
  \end{subequations}
   \negthinspace\negthinspace\negthinspace
   The number of contributing Feynman diagrams
   and the number of color bases for the above processes are summarized in
   Table~\ref{tab:qcd-processes}, which includes those for the process,
   $gg\!\rightarrow\! 5g$.
   We note here that the number of diagrams (7245) for $gg\!\rightarrow\! 5g$
   exceeds that of the $u\overline{u}\!\rightarrow\! 7\gamma$ process
   ($7!\!\approx\! 5040$), for which we could run the converted MadGraph
   codes on a GPU, only after division into small pieces~\cite{qed-paper}.
   In fact, we have not been able to run the $gg\!\rightarrow\! 5g$
   program on GPU even after dividing the program into more than 100
   pieces;  as explained in section~\ref{sec:gg5g}.
   
   Proliferation of the number of independent color basis vectors is also
   a serious concern for GPU computing, since the color matrix $\mathcal{N}$
   of eq.~(\ref{eq:color-matrix}) has $m(m+1)/2$ elements when there are
   $m$ independent basis vectors $T_{a}^{c_{i}}$.
   For example, the process $uu\!\rightarrow\! uuggg$ has $m\!=\!240$
   color basis 
   vectors from Table~\ref{tab:qcd-processes}, and the matrix has
   $3\!\times\! 10^{4}$ elements. 
   The matrix exceeding 16000 elements cannot be stored in the 64kB
   constant  memory, while storing it in the global memory will result in
   serious  loss of efficiency in parallel computing. 
   Therefore, the method to handle summation over color degrees of
   freedom is a serious concern in GPU computing.

  \subsection{Selection criteria for jets}
  \label{subsec:cuts}
  
   Total and differential cross sections of the
   processes~(\ref{eq:processes}) in $pp$ 
   collisions at $\sqrt{s}\!=\!14\textrm{TeV}$ are computed in this paper.
   We introduce final state cuts for all the jets as follows:
   \begin{subequations}
    \begin{eqnarray}
     |\eta_{i}| &<& \eta^{\scriptstyle\mathrm{cut}} = 2.5, \label{eq:cuts-eta}\\
     p_{\mathrm{T}i} &>& p_{\mathrm{T}}^{\scriptstyle\mathrm{cut}} =
     20\,\mathrm{GeV} , \label{eq:cuts-pt}\\
     p_{\mathrm{T}ij} &>& p_{\mathrm{T}}^{\scriptstyle\mathrm{cut}} =
      20\,\mathrm{GeV}, \label{eq:cuts-ptjj}
    \end{eqnarray}
    \label{eq:cuts}
   \end{subequations}
   \negthinspace\negthinspace\negthinspace
   where  $\eta_{i}$ and $p_{\mathrm{T}i}$ are the rapidity and the
   transverse momentum of the $i$-th jet, respectively, in the $pp$
   collisions rest frame along the right-moving ($p_{\mathrm{z}}\!=\!|p|$)
   proton momentum direction, and $p_{\mathrm{T}ij}$ is the relative
   transverse momentum~\cite{durham-jet} between the jets $i$ and $j$
   defined by  
   \begin{subequations}
    \begin{eqnarray}
     p_{\mathrm{T}ij}  \equiv
     \min(p_{\mathrm{T}i},p_{\mathrm{T}j})\,\Delta R_{ij}, \\
     \Delta R_{ij}  =
      \sqrt{\Delta\eta_{ij}^{2}+\Delta\phi_{ij}^{2}}.
    \end{eqnarray}
   \end{subequations}
   Here $\Delta R_{ij}$ measures the boost-invariant angular separation
   between the jets.
   
   As for the parton distribution function (PDF), we use the set
   CTEQ6L1~\cite{cteq}  and the factorization scale is chosen to be the
   cut-off $p_{\mathrm{T}}$  value,  
   $Q\! = p_{\mathrm{T}}^{\scriptstyle\mathrm{cut}}\! =\! 20\,\textrm{GeV}$.
   The QCD coupling constant is also fixed as
   \begin{equation}
    \alpha_{s} = 
    \alpha_{s}(Q\!=\!20\mathrm{GeV})_{\overline{\mathrm{MS}}} = 0.171\,,
   \end{equation}
   which is obtained from the
   $\overline{\mathrm{MS}}$ coupling at $Q\!=\!m_{Z}$,
   $\alpha_s(m_{Z})_{\overline{\mathrm{MS}}}\! =\!  0.118$~\cite{pdg} by
   using the NLO renormalization group equations with  5-flavors. 

 \section{Computation on the GPU}

 \subsection{GPU and its host PC}

  For the computation of the cross sections of QCD $n$-jet production
  processes we use the same GPU and host PC as in the previous 
  report~\cite{qed-paper}. 
  In particular we use a GeForce GTX280 by NVIDIA~\cite{nvidia} with 240
  processors, whose parameters are summarized in Table~\ref{tab:gpu}.
  It is controlled by a Linux PC with Fedora 8 on a CPU whose properties
  are summarized in Table~\ref{tab:host-pc}. 
  \begin{table}[tbh]
   \begin{center}
    \caption{Parameters of GeForce GTX280.}
    \label{tab:gpu}       
    \smallskip
    \begin{tabular}{|c|c|} \hline
     Number of & 30   \\
     multiprocessor & \\ \hline
     Number of core & 240   \\ \hline
     Total amount of  &	 1000  \\
     global memory [MB] &  \\ \hline
     Total amount of  &	 64   \\
     constant memory [kB] & \\ \hline
     Total amount of shared & 16  \\
     memory per block [kB] &  \\ \hline
     Total number of registers  & 16  \\
     available per block [kB] &  \\ \hline
     Clock rate [GHz] &	1.30  \\
     \hline
    \end{tabular}
   \end{center}
   \begin{center}
    \caption{Host PC environment.}
    \label{tab:host-pc}       
    \begin{tabular}{|c|c|} \hline
     CPU & Core2Duo 3GHz \\ \hline
     L2 Cache & 6MB  \\ \hline
     Memory & 4GB  \\ \hline
     Bus Speed & 1.333GHz  \\ \hline 
     OS & Fedora 8 (64 bit)  \\ \hline
    \end{tabular}
   \end{center}
  \end{table}
  
  Programs which are used for the computation of the cross sections are
  developed with the CUDA~\cite{cuda} environment introduced by
  NVIDIA~\cite{nvidia} for general purpose GPU computing. 

 \subsection{Program structure}

  Our program computes the total cross sections and distributions of the
  QCD $n$-jet production processes via the following procedure:
  \begin{enumerate}
   \item initialization of the program,
   \item random number generation for multiple phase-space points
	 $\{p_{a},p_{b},p_{1},\ldots,p_{n}\}$ and
	 helicities $\{\lambda_{i}\}$ on the
	 CPU, 
   \item transfer of random numbers to the GPU,
   \item generation of helicities and momenta of initial and final
	 partons using random numbers, and compute
	 amplitudes $(J_{\lambda_{i}})_{a}$ of
	 eq.~(\ref{eq:color-amplitude}) for all the color bases on
	 the GPU, 
   \item multiplying the amplitudes and their complex conjugate with the
	 color matrix $\mathcal{N}_{ab}$ of eq.~(\ref{eq:color-matrix})
	 and summing them up as in eq.~(\ref{eq:color-averaged-amplitude}),
	 and multiply the PDF's of the incoming partons on the GPU, 
   \item transferring momenta and helicities for external particles,
	 computed weights and the color summed squared amplitudes to the
	 CPU, and 
   \item summing up all values to obtain the total cross section and
	 distributions on the CPU.
  \end{enumerate}
  Program steps between the generation of random numbers~(2) and the
  summation of computed cross sections~(7) are repeated until we obtain
  sufficient statistics for the cross section and all distributions.

  \subsection{Color matrix calculation}
  \label{subsec:color-matrix}

  In order to compute the cross sections of the QCD multi-jet
  production processes, multiplications of the large color matrix
  $\mathcal{N}_{ab}$ of eq.~(\ref{eq:color-matrix}), the
  vector of color-bases amplitudes $(J_{\lambda{i}})_{\alpha}$ of
  eq.~(\ref{eq:color-amplitude}) and its complex conjugate have to be 
  performed, as in eq.~(\ref{eq:color-averaged-amplitude}).
  For large $n$-jet processes,  like $gg\!\rightarrow\! 4\,\mathrm{gluons}$,
  $u\overline{u}\!\rightarrow\! 5\,\mathrm{gluons}$ and $uu\!\rightarrow\!
  uu\!+\!3\,\mathrm{gluons}$, the dimensions of color matrices exceed 100,
  and the number of multiplication becomes larger than $10^{4}$.
  These matrices cannot be stored in the constant memory (64kB for
  the GTX280; see Table~\ref{tab:gpu}) which is
  accessed in parallel, while storing them in the global memory (1GB for
  GTX280) results in  serious slow-down of the GPU.
  We find that multiplications for the color-summation in
  eq.~(\ref{eq:color-averaged-amplitude}) can be reduced significantly
  as follows. 

  The color matrix of eq.~(\ref{eq:color-matrix}) contains many elements
  with the same value.
  We count the number of different non-zero elements in the color matrix
  and find the results shown in Table~\ref{tab:color-matrix}.
  We find for instance that among the $240\!\times\!(240\!+\!1)/2 =
  28,720$ elements of the color matrix for the $uu\!\rightarrow\! uu+3g$
  process, there are only 60 unique ones.

  In general, the number of different elements in the color matrix grows
  linearly rather than quadratically as the number of color basis
  vectors grows.
  Since the numbers in Table~\ref{tab:color-matrix} are small enough, we
  can store them in the constant memory which is accessed quickly by
  each parallel processor.

  Before we arrive at the above solution adopted in this study, we
  examined the possibility of summing over colors via Monte Carlo. 
  Let us briefly report, in passing, on this exercise.

  In the Monte Carlo color summation approach, we evaluate the matrix
  element $\mathcal{M}_{\lambda_{i}}^{c_{i}}$ (\ref{eq:matrix-element})
  for a given set of momenta $\{p_{i}\}$, helicities $\{\lambda_{i}\}$
  and colors $\{c_{i}\}$,  and sum the squared amplitudes over randomly 
  generated sets of $\{p_{i}, \lambda_{i}, c_{i}\}$.
  This method turns out not to be efficient because in the color basis
  using the fundamental representation of the SU(3) generators adopted
  by MadGraph, most of the basis vectors $T_{a}^{c_{i}}$ vanish for a
  given color configuration $\{c_{i}\}$.
  As an example, $gg\rightarrow 4g$ has $5!\!=\!120$ color basis vectors
  (see Table~\ref{tab:qcd-processes}), which take the form
  \begin{equation}
   T_{\alpha}^{c_{i}} = \mathrm{Tr}\,(T^{a_{1}}T^{a_{2}}\cdots
    T^{a_{6}}) 
    \label{eq:color-basis}
  \end{equation}
  for the configuration $\{c_{i}\}\! =\! (a_{1},a_{2},\ldots,a_{6})$ where
  $a_{i}$ denotes the color index of the gluon $i$ taking an integer
  value between 1 and 8.
  Among the $8^6\!\approx\! 260,000$ configurations, only 12\% give non-zero
  values.
  Moreover, as many as 75\% of the color configurations  give vanishing
  results for all the 120 basis vectors.
  Although the efficiency can be improved by changing the color basis,
  we find that our solution of evaluating the exact summation over
  colors is superior to the Monte Carlo summation method for all the
  processes which we report in this paper. 
   \begin{table}[tb]
    \begin{center}
     \caption{Number of different non-zero elements in the color matrix
     of eq.~(\ref{eq:color-matrix}).}
     \label{tab:color-matrix}       
     \begin{tabular}{|c|c|c|c|} \hline
      No. of jets
      & $gg\!\rightarrow\!\mathrm{gluons}$
      & $u\overline{u}\!\rightarrow\!\mathrm{gluons}$
      & $uu\!\rightarrow\! uu\!+\!\mathrm{gluons}$ \\
      \hline \hline
      2 & 3 & 2 & 2 \\ \hline
      3 & 7 & 4 & 7 \\ \hline
      4 & 15 & 9 & 19 \\ \hline
      5 & 45 & 24 & 60 \\ \hline
     \end{tabular}
    \end{center}
   \end{table}
  
  \section{New HEGET functions}
  \label{sec:heget}
  
  The HEGET functions for massless quarks and gluons are the same as
  those introduced in the previous report~\cite{qed-paper}.
  The $qqg$ vertex functions are identical to the $qq\gamma$ functions
  of ref.~\cite{qed-paper} except for the coupling constant;
  \begin{equation}
   eQ_{q} \rightarrow g_{s} T_{i\bar{j}}^{a}
  \end{equation}
  for the vertex
  \begin{equation}
   \mathcal{L}_{qqg} = -g_{s} T_{i\bar{j}}^{a} A_{\mu}^{a}(x)
    \overline{q}_{\bar{i}}(x)\gamma_{\mu}q_{j}(x)
    \label{eq:qqg-lagrangian}
  \end{equation}
  where $g_{s}\!=\!\sqrt{4\pi\alpha_{s}}$ is the strong coupling constant
  and $T_{i\bar{j}}^{a}$ is an SU(3)) generator in the fundamental
  representation. 
  For example, the $qqg$ vertex function is computed by the HEGET function
  \texttt{iovxx0} as
  \begin{equation} 
   \begin{array}{r}
   \texttt{iovxx0(cmplx* fi, cmplx* fo, cmplx* vc, float g,} \\
   \texttt{cmplx \pmb{vertex})} 
   \end{array}
  \end{equation}
  where the coupling constants are
  \begin{equation}
   \mathtt{g[0]} = \mathtt{g[1]} = g_{s}
  \end{equation}
  following the convention of MadGraph~\cite{madgraph}
  and the color amplitude is
  \footnotemark[1]
  \footnotetext[1]{The sign of the color amplitudes (\ref{eq:qqg-amplitude})
  and (\ref{eq:vvv-amplitude}) follows the sign of the Lagrangian 
  terms (\ref{eq:qqg-lagrangian}) and (\ref{eq:vvv-lagrangian}),
  respectively.
  MadGraph~\cite{madgraph} adopts the Lagrangian with the opposite sign,
  that is, $(g_{s})_{\mathrm{MadGraph}} = -g_{s}$.
  This sign difference is absorbed by the conventions
  (\ref{eq:qqg-amplitude}) and (\ref{eq:vvv-amplitude}).}
  \begin{equation}
   -T_{i\bar{j}}^{a}\, (\pmb{\texttt{vertex}})\,.
    \label{eq:qqg-amplitude}
  \end{equation}

  In the rest of this section, we introduce new HEGET functions for
  three-vector boson (\texttt{VVV}) and four-vector boson
  (\texttt{VVVV}) vertices.
  All the new HEGET functions are listed in Table~\ref{tab:vertices},
  and their contents are given in Appendix.
  Also shown in Table~\ref{tab:vertices} is the correspondence between
  the HEGET functions and the HELAS subroutines~\cite{helas}.
  \begin{table*}[htb]
   \begin{center}
    \caption{List of the new vertex functions in HEGET.}
    \label{tab:vertices}       
    \begin{tabular}{|c|c|c|c|c|} \hline
     Vertex & Inputs & Output & HEGET Function & HELAS Subroutine
     \\ \hline \hline
     VVV & VVV & Amplitude & {\ttfamily vvvxxx} & \texttt{VVVXXX} \\
         & VV  & V         & {\ttfamily jvvxx0} & \texttt{JVVXXX} \\ \hline
     VVVV & GGGG & Amplitude & {\ttfamily ggggxx} & \texttt{GGGGXX} \\
          & GGG  & G         & {\ttfamily jgggx0} & \texttt{JGGGXX} \\
     \hline
    \end{tabular}
   \end{center}
  \end{table*}

  \subsection{VVV: three vector boson vertex}

  For the $ggg$ vertex
  \begin{equation}
   \mathcal{L}_{ggg} = g_{s}f^{abc}
    ({\partial^{\mu}\!A^{a\nu}}(x))
    {A_{\mu}^{b}}(x){A_{\nu}^{c}}(x)
    \label{eq:vvv-lagrangian}
  \end{equation}
  we introduce two HEGET functions, \texttt{vvvxxx} and \texttt{jvvxx0}.
  They correspond to HELAS subroutines, \texttt{VVVXXX},
  and \newline\texttt{JVVXXX}, respectively, for massless particles; see
  Table~\ref{tab:vertices}. 
  
  \subsubsection{\texttt{vvvxxx}}

  The HEGET function \texttt{vvvxxx} (List~\ref{list:vvvxxx} in
  Appendix) computes the amplitude of 
  the \texttt{VVV} vertex from vector boson wave functions, whether they
  are on-shell or off-shell.
  The function has the arguments:
  \begin{equation} 
   \begin{array}{r}
   \texttt{vvvxxx(cmplx* ga, cmplx* gb, cmplx* gc, } \\
   \texttt{float g, cmplx \pmb{vertex})} 
    \end{array}
    \label{eq:vvv-function}
  \end{equation}
  where the inputs and the outputs are:
  \begin{equation}
   \begin{array}{ll}
    \textsc{Inputs:} & \\
    \texttt{cmplx ga[6]} & \textrm{wavefunction of gluon with color} \\
    & \textrm{index, $a$} \\ 
    \texttt{cmplx gb[6]} & \textrm{wavefunction of gluon with color} \\
    & \textrm{index, $b$} \\ 
    \texttt{cmplx gc[6]} & \textrm{wavefunction of gluon with color} \\
    & \textrm{index, $c$} \\ 
     \texttt{float g} & \textrm{coupling constant of \texttt{VVV}
      vertex} \\ \\
    \textsc{Outputs:} & \\
    \texttt{cmplx vertex} & \textrm{amplitude of the \texttt{VVV} 
     vertex} 
     \label{eq:vvv}
    \end{array}
  \end{equation}
  The coupling constant is
  \begin{equation}
   \mathtt{g} = g_{s}
    \label{eq:vvv-coupling}
  \end{equation}
  in the HEGET function~(\ref{eq:vvv-function}), following the
  convention of MadGraph~\cite{madgraph}.
  In order to reproduce the amplitudes associated
  with the $ggg$ vertex Lagrangian of eq.~(\ref{eq:vvv-lagrangian}), the
  color factor 
  associated with the $ggg$ vertex is $i f^{abc}$.
  More explicitly, the vertex amplitude for
  eq.~(\ref{eq:vvv-lagrangian}) is\footnotemark[1]
  \begin{equation}
   i f^{abc}\,(\pmb{\texttt{vertex}})
    \label{eq:vvv-amplitude}
  \end{equation}
  by using the output (\pmb{\texttt{vertex}}) in
  eq.~(\ref{eq:vvv-function}).   
  Also note the HELAS convention~\cite{helas} of using the flowing-OUT
  momenta and quantum numbers for all bosons. 

  \subsubsection{\texttt{jvvxx0}}

  This HEGET function \texttt{jvvxx0} (List~\ref{list:jvvxx0} in
  Appendix) 
  computes the off-shell vector wavefunction from the
  three-point gauge boson coupling in eq.~(\ref{eq:vvv-lagrangian}).  
  The vector propagator is given in the Feynman gauge for a massless
  vector bosons like gluons.
  It has the arguments:
  \begin{equation} 
   \begin{array}{r}
   \texttt{jvvxx0(cmplx* ga, cmplx* gb, float g,} \\
   \texttt{cmplx* \pmb{jvv})} 
   \end{array}
  \end{equation}
  where the inputs and the outputs are:
  \begin{equation}
   \begin{array}{ll}
    \textsc{Inputs:} & \\
    \texttt{cmplx ga[6]} & \textrm{wavefunction of gluon with color} \\
    & \textrm{index, $a$} \\ 
    \texttt{cmplx gb[6]} & \textrm{wavefunction of gluon with color} \\
    & \textrm{index, $b$} \\ 
     \texttt{float g} & \textrm{coupling constant of the
      \texttt{VVV} vertex}  \\ \\
    \textsc{Outputs:} & \\
     \texttt{cmplx jvv[6]} & \textrm{vector current
      $j^{\mu}(\texttt{gc}\!:\!\texttt{ga},\texttt{gb})$ which has} \\
    & \textrm{a color index, $c$}
    \end{array}
  \end{equation}
  As in eq.~(\ref{eq:vvv-amplitude}) the color amplitude for the
  off-shell current is 
  \begin{equation}
   i f^{abc}\,(\pmb{\texttt{jvv}})\,.
  \end{equation}

  \subsection{VVVV: four vector boson vertex}

  For the $ggggg$ vertex
  \begin{equation}
   \mathcal{L}_{gggg} = - \frac{g_{s}^2}{4}
    f^{abe}f^{cde}{A^{a\mu}}(x){A^{b\nu}}(x)
    {A_{\mu}^{c}}(x){A_{\nu}^{d}}(x)
  \end{equation}
  we introduce two HEGET functions, \texttt{ggggxx} and \texttt{jgggx0}, 
  listed in Table~\ref{tab:vertices}.
  They correspond to HELAS subroutines, \texttt{GGGGXX}
  and \texttt{JGGGXX}, respectively, for massless particles.

  \subsubsection{\texttt{ggggxx}}

  The HEGET function \texttt{ggggxx} (List~\ref{list:ggggxx} in
  Appendix) computes the portion of the amplitude of the $gggg$
  amplitude where the first and the third, and hence also, the second
  and the fourth gluon wave functions are contracted, whether the gluons
  are  on-shell or off-shell.
  The function has the arguments:
  \begin{equation} 
   \begin{array}{r}
   \texttt{ggggxx(cmplx* ga, cmplx* gb, cmplx* gc,} \\
   \texttt{cmplx* gd, float g, cmplx vertex)} 
    \end{array}
  \end{equation}
  where the inputs and the outputs are:
  \begin{equation}
   \begin{array}{ll}
    \textsc{Inputs:} & \\
     \texttt{cmplx ga[6]} & \textrm{wavefunction of gluon with 
      color} \\
    & \textrm{index, $a$} \\ 
     \texttt{cmplx gb[6]} & \textrm{wavefunction of gluon with 
      color} \\
    & \textrm{index, $b$} \\ 
     \texttt{cmplx gc[6]} & \textrm{wavefunction of gluon with 
      color} \\
    & \textrm{index, $c$} \\ 
     \texttt{cmplx gd[6]} & \textrm{wavefunction of gluon with 
      color} \\
    & \textrm{index, $d$} \\ 
     \texttt{float gg} & \textrm{coupling constant of \texttt{VVV}
      vertex} \\ \\
    \textsc{Outputs:} & \\
    \texttt{cmplx vertex} & \textrm{amplitude of the \texttt{VVVV} 
     vertex} 
    \end{array}
  \end{equation}

  \smallskip\noindent
  The coupling constant \texttt{gg} for the $gggg$ vertex is
  \begin{equation}
   \texttt{gg} = g_{s}^{2}\,.
  \end{equation}
  In order to obtain the complete amplitude, the function must be called
  three times (once for each color structure) with the following
  permutations:
  \begin{subequations}
   \begin{eqnarray}
    \mathtt{ggggxx(ga,gb,gc,gd,gg,\pmb{\texttt{v1}})} & \\
    \mathtt{ggggxx(ga,gc,gd,gb,gg,\pmb{\texttt{v2}})} & \\
    \mathtt{ggggxx(ga,gd,gb,gc,gg,\pmb{\texttt{v3}})} &
   \end{eqnarray}
   \label{eq:gggg-functions-calls}
  \end{subequations}
  The color amplitudes are then expressed as
  \begin{equation}
   f^{abe} f^{cde}\, (\pmb{\texttt{v1}})
    + f^{ace} f^{dbe}\, (\pmb{\texttt{v2}})
    + f^{ade} f^{bce}\, (\pmb{\texttt{v3}})\,.
  \end{equation}
  
  \subsubsection{\texttt{jgggx0}}

  The HEGET function \texttt{jgggx0} (List~\ref{list:jgggx0} in Appendix)
  computes an off-shell gluon current from the 
  four-point gluon coupling, including the gluon propagator in the
  Feynman gauge. 
  It has the arguments:
  \begin{equation} 
   \begin{array}{r}
   \texttt{jgggx0(cmplx* ga, cmplx* gb, cmplx* gc, float gg,} \\
   \texttt{cmplx* jggg)} 
   \end{array}
   \label{eq:jggg-func}
  \end{equation}
  where the inputs and the outputs are:
  \begin{equation}
   \begin{array}{ll}
    \textsc{Inputs:} & \\
     \texttt{cmplx ga[6]} & \textrm{wavefunction of gluon with
      color} \\ 
    & \textrm{index, $a$} \\ 
     \texttt{cmplx gb[6]} & \textrm{wavefunction of gluon with
      color} \\
    & \textrm{index, $b$} \\ 
     \texttt{cmplx gc[6]} & \textrm{wavefunction of gluon with
      color} \\
    & \textrm{index, $c$} \\ 
     \texttt{float gg} & \textrm{coupling constants of the
      \texttt{VVVV} vertex}  \\ \\
    \textsc{Outputs:} & \\
     \texttt{cmplx jggg[6]} & \textrm{vector current
      $j^{\mu}(\texttt{gd}\!:\!\texttt{ga},\texttt{gb},\texttt{gc})$
      which} \\ 
    & \textrm{has a color index, $d$} 
    \end{array}
  \end{equation}
  The function (\ref{eq:jggg-func}) computes off-shell gluon wave
  function with three specific color index $d$ which comes along with
  a specific color factor. 
  As in eq.~(\ref{eq:gggg-functions-calls}) it should be called three
  times
  \begin{subequations}
   \begin{eqnarray}
    \texttt{jgggx0(ga,gb,gc,gg,\pmb{j1})} & \\
    \texttt{jgggx0(gc,ga,gb,gg,\pmb{j2})} & \\
    \texttt{jgggx0(gb,gc,ga,gg,\pmb{j3})} &
   \end{eqnarray}
   \label{eq:ggg-functions}
  \end{subequations}
  to give the off-shell gluon with the color factor
  \begin{equation}
   f^{abe} f^{cde}\, (\pmb{\texttt{j1}})
    + f^{ace} f^{dbe}\, (\pmb{\texttt{j2}})
    + f^{ade} f^{bce}\, (\pmb{\texttt{j3}})\,.
  \end{equation}
  
 \section{Results}
 \label{sec:results}
 
  \subsection{Comparison of total cross sections}
  
   In order to validate the new HEGET functions which are introduced in
   this report,  we compare the total cross 
   sections of $n$-jet production processes computed on the GPU with
   those calculated by other programs which are based on 
   the FORTRAN version of the HELAS library.
   We use MadGraph/MadEvent~\cite{madgraph} and another independent 
   FORTRAN program which uses the Monte
   Carlo integration program, BASES~\cite{bases}, as references.
   Due to the limited support for the double precision computation
   capabilities on the GPU, the whole computations with HEGET on a
   GTX280 are done with single precision, while the other programs with
   HELAS in FORTRAN compute cross sections with double precision.

   For the calculation of the $n$-jet production cross sections we use
   the same physics parameters as the MadGraph/MadEvent for all
   programs, and the same final state cuts of eq.~(\ref{eq:cuts}) for all
   processes and all programs.
   The parton distribution
   functions of CTEQ6L1~\cite{cteq} and the same factorization and
   renormalization scales, $Q =
   p_{\mathrm{T}}^{\scriptstyle\mathrm{cut}} = 20\textrm{GeV}$, are
   also used.  

   Results for the computation of the total cross sections are
   summarized in Tables \ref{tab:sigma-ggng}, \ref{tab:sigma-uuxng} and  
   \ref{tab:sigma-uuuung} for $gg\!\rightarrow\! \mathrm{gluons}$,
   $u\overline{u}\!\rightarrow\! \mathrm{gluons}$ and $uu\!\rightarrow\!
   uu\!+\!\mathrm{gluons}$, respectively.
   We find the results obtained by the HEGET
   functions agree with those from the other programs within the statistics
   of generated number of events.

   We note that multi-jet events that satisfy the final state cuts
   of eq.~(\ref{eq:cuts}), where all jets are in the central region in
   $|\eta|\!<\!2.5$ (\ref{eq:cuts-eta}) and their transverse momentum
   about the beam direction (\ref{eq:cuts-pt}) and among each other
   (\ref{eq:cuts-ptjj}) greater than 20 GeV, are dominated by pure
   gluonic processes in Table~\ref{tab:sigma-ggng}.
   The cross sections for $u\overline{u}\!\rightarrow\! ng$ process in
   Table~\ref{tab:sigma-uuxng} are small because of $u\overline{u}$
   annihilation.
   We note that the crossing-related non-annihilation processes,
   $ug\!\rightarrow\! u\!+\!(n\!-\!1)g$, have exactly the same number of
   diagrams and color bases, hence can be evaluated with
   essentially the same amount of computation time.

  \begin{table*}[htb]
   \begin{center}
    \caption{Total cross sections for $gg\!\rightarrow\!\mathrm{gluons}$ [fb].}
    \label{tab:sigma-ggng}       
    \begin{tabular}{ccccl} \hline
     No. of jets
     & HEGET & Bases & MadGraph/MadEvent & \\ \hline
     \noalign{\smallskip}
     2 & $3.1929 \pm 0.0010$  
       & $3.1928 \pm 0.0010$ 
       & $3.1902 \pm 0.0076$ 
       & $\times 10^{11}$ \\
     3 & $2.6201 \pm 0.0023$  
       & $2.6136 \pm 0.0036$ 
       & $2.6221 \pm 0.0061$ 
       & $\times 10^{10}$ \\
     4 & $5.813 \pm 0.020$ 
       & $5.8140\pm 0.0095$ 
       & $5.776\pm 0.034$ 
       & $\times 10^{9}$ \\
     \noalign{\smallskip}\hline
    \end{tabular}
   \end{center}
   \begin{center}
    \caption{Total cross sections for
    $u\overline{u}\!\rightarrow\!\mathrm{gluons}$ [fb].} 
    \label{tab:sigma-uuxng}       
    \begin{tabular}{ccccl} \hline
     No. of jets
     & HEGET & Bases & MadGraph/MadEvent & \\ \hline
     \noalign{\smallskip}
     2 & $2.8981 \pm 0.0007$  
       & $2.8969 \pm 0.0006$ 
       & $2.8991 \pm 0.0073$ 
       & $\times 10^{7}$ \\
     3 & $1.8420 \pm 0.0012$ 
       & $1.8388 \pm 0.0018$ 
       & $1.8421 \pm 0.0077$ 
       & $\times 10^{6}$ \\
     4 & $4.465 \pm 0.022$ 
       & $4.496 \pm 0.017$ 
       & $4.399 \pm 0.038$ 
       & $\times 10^5$ \\
     5 & $1.566 \pm 0.057$ 
       & $1.589 \pm 0.018$ 
       & $1.542 \pm 0.039$ 
       & $\times 10^{5}$ \\
     \noalign{\smallskip}\hline
    \end{tabular}
   \end{center}
   \begin{center}
    \caption{Total cross sections for $uu\!\rightarrow\!
    uu\!+\!\mathrm{gluons}$ [fb].} 
    \label{tab:sigma-uuuung}       
    \begin{tabular}{ccccl} \hline
     No. of jets
     & HEGET & Bases & MadGraph/MadEvent & \\ \hline
     \noalign{\smallskip}
     2 & $2.6715 \pm 0.0014$  
       & $2.6743 \pm 0.0011$ 
       & $2.6689 \pm 0.0047$ 
       & $\times 10^{8}$ \\
     3 & $5.897 \pm 0.004 $  
       & $5.889 \pm 0.010$ 
       & $5.871 \pm 0.015$ 
       & $\times 10^{7}$ \\
     4 & $2.7754 \pm 0.0130$ 
       & $2.7500 \pm 0.0083$ 
       & $2.748 \pm 0.042$ 
       & $\times 10^{7}$ \\
     5 & $1.513 \pm 0.024$   
       & $1.560 \pm 0.013$ 
       & $1.513 \pm 0.024$ 
       & $\times 10^{6}$ \\
     \noalign{\smallskip}\hline
    \end{tabular}
   \end{center}
  \end{table*}
  
  \subsection{Comparison of the processing time}

  As already described in our previous report~\cite{qed-paper}, we
  prepare two versions of the programs in the same structure for the
  computation of the total cross sections.  One is written in CUDA,
  a C-based language, and can be executed on the GPU.  The other
  is written in C and can be executed on the CPU.
  Using a standard C library function we measure the time between the
  start of the transfer of random  numbers to the GPU and the end of the
  transfer of computed results back to the CPU.

\begin{figure}[htb]
   \begin{center}
    \resizebox{0.45\textwidth}{!}{%
    \includegraphics{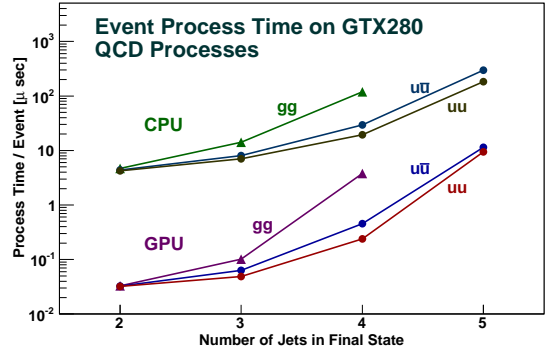}
    }
    \caption{Processing time for GPU and CPU.}
    \label{fig:time}       
   \end{center}
  \end{figure}

  In Fig.~\ref{fig:time}, the measured process time in $\mu$sec for one
  event of  $n$-jet production processes is shown for the GPU (GTX280)
  and the CPU  (Linux PC with Fedora 8). 
  They are plotted against the number of jets in the final state.
  Because the process time per event on the GPU depends~\cite{qed-paper}
  strongly on the number of allocated registers at the 
  compilation by the CUDA and the size of thread blocks at the execution
  time, we scan combination of these parameters for the fastest event
  process time on the GPU.

  The upper three lines in Fig.~\ref{fig:time} show the event process
  times on the CPU.
  They correspond to $gg\!\rightarrow\! n$-jets denoted as $\mathbf{gg}$,
  $u\overline{u}\!\rightarrow\! n$-jets as
  $\mathbf{u\overline{u}}$ and $uu\!\rightarrow\! uu\!+\!n$-jets as
  $\mathbf{uu}$, respectively. 
  For processes with small numbers of jets, e.g. $n_\textrm{jet}\!=\!2$,
  the event process times for different processes are all
  around 4.5~$\mu$sec.
  This is probably because they are dominated
  by computation steps other than the amplitude calculations,
  such as computations of the PDF factors and the data transfer between
  GPU  and CPU, which are common to all physics processes. 
  When the number of jets becomes larger, the event process time for the
  same number jets in the final states is roughly proportional to the
  number of diagrams of each process listed in
  Table~\ref{tab:qcd-processes}. 

  The lower three lines in Fig.~\ref{fig:time} show the event process
  times on a GTX280.
  They also correspond to $gg\!\rightarrow\! n$-jets denoted as
  $\mathbf{gg}$,  $u\overline{u}\!\rightarrow\!
  n$-jets as $\mathbf{u\overline{u}}$ and $uu\!\rightarrow\!
  uu\!+\!n$-jets as $\mathbf{uu}$, respectively.
  As the number of jets becomes larger, the process time on the GPU
  grows more rapidly than that on the CPU.
  For the $n_{\textrm{jet}}\!=\!4$ case, the event process time of
  $gg\!\rightarrow\! 4\!\textrm{ gluons}$ is larger than the expected
  time from the proportionality to the number of diagrams of the other
  processes, $u\overline{u}\!\rightarrow\! 4\!\textrm{ gluons}$ and
  $uu\!\rightarrow\! uu\!+\!2\!\textrm{ gluons}$. 
  In other words, the event process time on GPU grows faster than what
  we expect from the growth of the number of Feynman diagrams.

  For instance, the event process times ratio for $gg\!\rightarrow\! 4g$
  and $gg\!\rightarrow\! 3g$ on the CPU are roughly 120~$\mu$sec/14~$\mu$sec 
  $\!\sim\! 8.6$, which roughly agrees with the ratio of the numbers of
  Feynman diagrams (Table~\ref{tab:qcd-processes}), $510/45\!\sim\! 11$.
  The corresponding ratio on GPU is 3.8~$\mu$sec/0.1~$\mu$sec $\!\sim\! 38$,
  which is significantly larger.

  For the same number of jets, we also observe that the event process
  times on the CPU are roughly proportional to the number of diagrams.
  Fir $n_{\mathrm{jet}}\!=\!4$, the ratio of the process times for
  $gg\!\rightarrow\! 4g$ to $u\overline{u}\!\rightarrow\! 4g$
  are about 120~$\mu$sec/29~$\mu$sec $\!\sim\! 4.1$ on CPU, as compared to
  the ratio of the number of Feynman diagrams in
  Table~\ref{tab:qcd-processes}, 
  $510/159\!\sim\! 3.2$.
  The same applies to $n_{\mathrm{jet}}\! = \!5$ between
  $u\overline{u}\!\rightarrow\! 5g$ and $uu\!\rightarrow\! uuggg$, where
  Feynman diagrams have the ratio $1890/786\!\sim\! 2.4$ from
  Table~\ref{tab:qcd-processes}, and the event process time on the CPU gives
  300~$\mu$sec/180~$\mu$sec $\!\sim\! 1.7$, also in rough
  agreement.

  On the other hand, the event process times on the GPU for $gg\!\rightarrow\!
  4g$ and $u\overline{u}\!\rightarrow\!4g$ have a ratio
  3.8~$\mu$sec/0.45~$\mu$sec $\!\sim\! 8.4$ which is much larger than the
  ratio of the diagram numbers; while that for
  $u\overline{u}\!\rightarrow\! 5g$ and $uu\!\rightarrow\! uuggg$ has
  the ratio of 11$\mu$sec/9.5$\mu$sec $\!\sim\! 1.15$. 
  Although we do not fully understand the above behavior of the event
  process time on the GPU, we find that they tends to scale as the 
  product of the number of Feynman diagrams and the number of color
  bases, while the event process times on the CPU are not sensitive to the
  latter. 
  This is probably because as the number of color bases grows, more
  amplitudes, $(J_{\lambda i})_{\alpha}$ in
  eq.~(\ref{eq:color-amplitude}), should be stored and then called to
  compute the color sum, eq.~(\ref{eq:color-averaged-amplitude}).
  These observations tell us that the relative weight of the color matrix
  computation in the GPU computing is very significant even after
  identifying the independent elements of the color matrix
  $\mathcal{N}_{\alpha\beta}$ in eq.~(\ref{eq:color-matrix}) as listed
  in Table~\ref{tab:color-matrix}.

  \subsection{Comparison of performance of GPU and CPU}
  
\begin{figure}[htb]
   \begin{center}
    \resizebox{0.45\textwidth}{!}{%
    \includegraphics{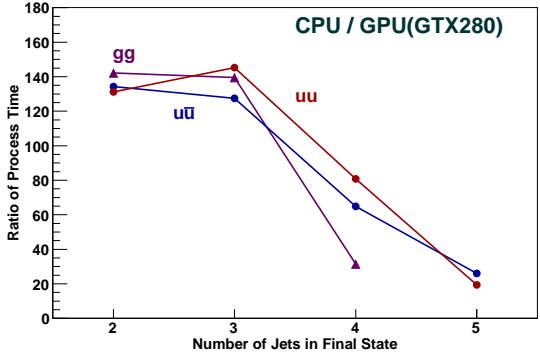}
    }
    \caption{Ratio of processing time. Time on CPU divided by time on GPU.}
    \label{fig:ratio}       
   \end{center}
  \end{figure}

  The ratios of event process times between CPU and GPU are shown in
  Fig.~\ref{fig:ratio}.
  Three lines correspond to $gg\!\rightarrow\! n$-jets denoted as
  $\mathbf{gg}$,  $u\overline{u}\!\rightarrow\!
  n$-jets as $\mathbf{u\overline{u}}$ and $uu\!\rightarrow\!
  uu\!+\!(n\!-\!2)$-jets as $\mathbf{uu}$, respectively.
  The performance ratios exceed 100 for the processes with small
  numbers of jets ($n_{\textrm{jet}}\!\leq\! 3$) in the final state.
  For $n_{\textrm{jet}}\!=\!4$ and 5, the performance ratios gradually
  drop to less than 40.
  For processes with large numbers of color bases, the ratios
  are smaller.
  For $gg\!\rightarrow\! 4\!\textrm{ gluons}$, which has 120 color bases,
  the ratio is about 30, and for $uu\!\rightarrow\! uu\!+\!3\!\textrm{
  gluons}$, which has 240 color bases, the ratio becomes about 20. 
  
  \subsection{Note on $gg\rightarrow 5g$ study}
  \label{sec:gg5g}
  
   Among five-jet production processes we have not been able to run the
   program for $gg\!\rightarrow\! 5g$.
   This process has 7245 diagrams and 720 color basis vectors.
   In order to compile the program for the computation of this
   process,  we use the technique developed in the previous
   study~\cite{qed-paper}. 
   By dividing the program into about 140 pieces we were able to
   compile the $gg\!\rightarrow\! 5g$ program.
   Compilation takes about 90~min. on a Linux PC.
   The total size of the compiled program exceeds 200~MB, and we
   were not able to execute this compiled program on a GTX280. 
   
   
 \section{Summary}
   \label{sec:summary}
   
   We have shown the results of our attempt to evaluate QCD
   multi-jet production processes at hadron colliders on a
   GPU~\cite{nvidia}, Graphic Processing Unit, following the
   encouraging results obtained for QED multi-photon production
   processes in ref.~\cite{qed-paper}. 
   
   Our achievements and findings may be summarized as follows.
   \begin{itemize}
    \item A new set of HEGET functions written in CUDA~\cite{cuda},
	  a C-language platform developed by NVIDIA for general purpose
	  GPU computing, are 
	  introduced  to compute triple and quartic gluon vertices.
	  The HEGET routines for massless quarks were introduced in
	  ref.~\cite{qed-paper},  and the routine for photons~\cite{qed-paper}
	  can be used for gluons.
	  In addition, the HEGET functions for the qqg vertex are the
	  same as those for the $qq\gamma$ vertex introduced in
	  ref.~\cite{qed-paper}. 
    \item The HELAS amplitude code generated by MadGraph~\cite{madgraph}
	  is converted to a CUDA program which calls HEGET functions
	  for the following three type of subprocesses:
	  $gg\!\rightarrow\! ng$ ($n\!\leq\! 5$),
	  $u\overline{u}\!\rightarrow\! ng$ ($n\!\leq\! 5$), and
	  $uu\!\rightarrow\! uu\!+\! ng$ ($n\!\leq\! 3$).
    \item Summation over color degrees of freedom was performed
	  on a GPU by identifying the same valued elements of the color
	  matrix of eq.~(\ref{eq:color-matrix}), in order to reduce the
	  memory size. 
    \item All the HEGET programs for up to 5 jets passed the CUDA compiler
	  after division into small pieces.
	  However, we could not execute the program for the process
	  $gg\!\rightarrow\! 5g$. 
	  Accordingly, comparisons of performance between GPU and
	  CPU are done for the multi-jet production processes up to 5
	  jets, excluding the purely gluonic subprocess.
    \item Event process times of the GPU program on GTX280 are more
	  than 100 times faster than the CPU program for all the
	  processes up to 3-jets, while the gain is reduced to 60
	  for 4-jets with one or two quark lines, and to 30 for
	  the purely gluonic process.
	  It further goes down to 30 and 20 for 5-jet production
	  processes with one and two quark lines, respectively.
    \item We find that one cause of the rapid loss of
	  GPU gain over CPU as the number of jets increases is the
	  growth in the number of color bases.
	  GPU programs slow down for processes with larger numbers of
	  color basis vectors, while the performance of the CPU programs
	  is not affected much.
    \item All computations on the GPU were performed with single
	  precision accuracy.
	  A factor of 2.5 to 4 slower performance is found for
	  double precision computation on the GPU.
   \end{itemize}

   \begin{acknowledgement}{\textit{Acknowledgement}.}
  We thank Johan Alwall, Qiang Li and Fabio Maltoni for stimulating
  discussions. 
  This work is supported  by  the Grant-in-Aid for Scientific
  Research from the Japan Society 
  for the Promotion of Science (No. 20340064) and
  the National Science Foundation (No. 0757889).
   \end{acknowledgement}
 
 \appendix
 \def\thesection{Appendix \Alph{section}}
 
 \section{Additional HEGET functions}

 In the appendix, we list the HEGET functions introduced in this 
 report.
 They are for the $ggg$ and $gggg$ vertices which do not have
 counterparts in QED.
 Together with the HEGET functions listed in ref.~\cite{qed-paper}, the
 quark and gluon (photon) wave functions and the $qqg (qq\gamma)$ vertices,
 all the QCD amplitudes can be computed on GPU.

 \subsection{Functions for the \texttt{VVV} vertex}

\begin{lstlisting}[caption=vvvxxx.cu,label=list:vvvxxx]
#include "cmplx.h"

__device__
void vvvxxx(cmplx* ga, cmplx* gb, cmplx* gc,
                           float  g, cmplx& vertex)
{
    cmplx v12 = ga[0]*gb[0]
        - ga[1]*gb[1] - ga[2]*gb[2] - ga[3]*gb[3];
    cmplx v23 = gb[0]*gc[0]
        - gb[1]*gc[1] - gb[2]*gc[2] - gb[3]*gc[3];
    cmplx v31 = gc[0]*ga[0]
        - gc[1]*ga[1] - gc[2]*ga[2] - gc[3]*ga[3];

    float pga[4];
    float pgb[4];
    float pgc[4];

    pga[0] = ga[4].re;
    pga[1] = ga[5].re;
    pga[2] = ga[5].im;
    pga[3] = ga[4].im;

    pgb[0] = gb[4].re;
    pgb[1] = gb[5].re;
    pgb[2] = gb[5].im;
    pgb[3] = gb[4].im;

    pgc[0] = gc[4].re;
    pgc[1] = gc[5].re;
    pgc[2] = gc[5].im;
    pgc[3] = gc[4].im;

    cmplx p12 = pga[0]*gb[0]
        - pga[1]*gb[1] - pga[2]*gb[2] - pga[3]*gb[3];
    cmplx p13 = pga[0]*gc[0]
        - pga[1]*gc[1] - pga[2]*gc[2] - pga[3]*gc[3];
    cmplx p21 = pgb[0]*ga[0]
        - pgb[1]*ga[1] - pgb[2]*ga[2] - pgb[3]*ga[3];
    cmplx p23 = pgb[0]*gc[0]
        - pgb[1]*gc[1] - pgb[2]*gc[2] - pgb[3]*gc[3];
    cmplx p31 = pgc[0]*ga[0]
        - pgc[1]*ga[1] - pgc[2]*ga[2] - pgc[3]*ga[3];
    cmplx p32 = pgc[0]*gb[0]
        - pgc[1]*gb[1] - pgc[2]*gb[2] - pgc[3]*gb[3];

    vertex = -(v12*(p13-p23)
               + v23*(p21-p31)
               + v31*(p32-p12))*g;

    return;
}
\end{lstlisting}
 
\begin{lstlisting}[caption=jvvxx0.cu,label=list:jvvxx0]
#include "cmplx.h"

__device__
void jggxxx(cmplx* ga, cmplx* gb, float g,
            cmplx* jvv)
{
    jvv[4] = ga[4] + gb[4];
    jvv[5] = ga[5] + gb[5];

    float p1[4];
    float p2[4];
    float q[4];

    p1[0] = (ga[4].re);
    p1[1] = (ga[5].re);
    p1[2] = (ga[5].im);
    p1[3] = (ga[4].im);

    p2[0] = (gb[4].re);
    p2[1] = (gb[5].re);
    p2[2] = (gb[5].im);
    p2[3] = (gb[4].im);

    q[0] = -(jvv[4].re);
    q[1] = -(jvv[5].re);
    q[2] = -(jvv[5].im);
    q[3] = -(jvv[4].im);

    float s   = q[0]*q[0]-q[1]*q[1]-q[2]*q[2]-q[3]*q[3];

    cmplx gab = ga[0]*gb[0]
        - ga[1]*gb[1] - ga[2]*gb[2] - ga[3]*gb[3];

    cmplx sga = (p2[0]-q[0])*ga[0] - (p2[1]-q[1])*ga[1]
        - (p2[2]-q[2])*ga[2] - (p2[3]-q[3])*ga[3];

    cmplx sgb = -(p1[0]-q[0])*gb[0] + (p1[1]-q[1])*gb[1]
        + (p1[2]-q[2])*gb[2] + (p1[3]-q[3])*gb[3];

    float gs = -g*(1.0f/s);

    jvv[0] = gs*((p1[0]-p2[0])*gab 
                 + sga*gb[0] + sgb*ga[0]);
    jvv[1] = gs*((p1[1]-p2[1])*gab 
                 + sga*gb[1] + sgb*ga[1]);
    jvv[2] = gs*((p1[2]-p2[2])*gab 
                 + sga*gb[2] + sgb*ga[2]);
    jvv[3] = gs*((p1[3]-p2[3])*gab 
                 + sga*gb[3] + sgb*ga[3]);

    return;
}
\end{lstlisting}

 \subsection{Functions for the \texttt{VVVV} vertex}
 
\begin{lstlisting}[caption=ggggxx.cu,label=list:ggggxx]
#include "cmplx.h"

__device__
void ggggxx(cmplx* ga, cmplx* gb, cmplx* gc,
            cmplx* gd, float gg, cmplx& vertex)
{

    cmplx gad = ga[0]*gd[0]
        -ga[1]*gd[1]-ga[2]*gd[2]-ga[3]*gd[3];

    cmplx gbc = gb[0]*gc[0]
        -gb[1]*gc[1]-gb[2]*gc[2]-gb[3]*gc[3];

    cmplx gac = ga[0]*gc[0]
        -ga[1]*gc[1]-ga[2]*gc[2]-ga[3]*gc[3];

    cmplx gbd = gb[0]*gd[0]
        -gb[1]*gd[1]-gb[2]*gd[2]-gb[3]*gd[3];

    vertex = gg*(gad*gbc-gac*gbd);

    return;
}
\end{lstlisting}
 
\begin{lstlisting}[caption=jgggx0.cu,label=list:jgggx0]
#include "cmplx.h"

__device__
void jgggx0(cmplx* ga, cmplx* gb, cmplx* gc, 
                         float gg, cmplx* jggg)
{

    jggg[4] = ga[4]+gb[4]+gc[4];
    jggg[5] = ga[5]+gb[5]+gc[5];

    float q[4];
    q[0] = -jggg[4].re;
    q[1] = -jggg[5].re;
    q[2] = -jggg[5].im;
    q[3] = -jggg[4].im;

    float fact = gg*(1.0f/(q[0]*q[0] 
                          -q[1]*q[1]-q[2]*q[2]-q[3]*q[3]));
    
    cmplx gcb = gc[0]*gb[0]
        -gc[1]*gb[1]-gc[2]*gb[2]-gc[3]*gb[3];
    cmplx gac = ga[0]*gc[0]
        -ga[1]*gc[1]-ga[2]*gc[2]-ga[3]*gc[3];

    jggg[0] = fact*( ga[0]*gcb - gb[0]*gac );
    jggg[1] = fact*( ga[1]*gcb - gb[1]*gac );
    jggg[2] = fact*( ga[2]*gcb - gb[2]*gac );
    jggg[3] = fact*( ga[3]*gcb - gb[3]*gac );


    return;
}

\end{lstlisting}

  %

\begin{thebibliography}{}
   %
   %
   
   \bibitem{qed-paper}
	   K.~Hagiwara, J.~Kanzaki, N.~Okamura, D.~Rainwater and T.~Stelzer,
	   {\ttfamily arXiv:0908.4403}.
   \bibitem{cuda}
	   \texttt{http://www.nvidia.com/object/cuda\_home.html}
   \bibitem{helas}
	   K.~Hagiwara, H.~Murayama and I.~Watanabe,
	   Nucl. Phys. \textbf{B367} (1991) 257;
	   H.~Murayama, I.~Watanabe and K.~Hagiwara,
	   KEK-Report 91-11, 1992.
   \bibitem{madgraph}
	   T.~Stelzer and W.~F.~Long,
	   Comput.\ Phys.\ Commun.\ {\bfseries 81} (1994) 357.
   \bibitem{su3}
	   See e.g.\ M.L.~Mangano and S.J.~Parke,
	   Phys.\ Rept.\ {\bfseries 200}, 301 (1991).
   \bibitem{durham-jet}
	   S.~Catani, Y.L.~Dokshitzer, M.H.~Seymour, B.R.~Webber,
	   Nucl.\ Phys.\ {\bfseries B406}, 187 (1993)
   \bibitem{cteq}
	   CTEQ Collaboration, H.L.~Lai et al.,
	   Eur.\ Phys.\ J. {\bfseries C12} (2000) 375.
   \bibitem{pdg}
	   C. Amsler et al. (Particle Data Group),
	   Phys.\ Lett.\ {\bfseries B667}, 1 (2008) and 
	   2009 partial update for the 2010 edition.
   \bibitem{madevent}
	   F.~Maltoni and T.~Stelzer,
	   JHEP {\bfseries 0302} (2003) 027.
   \bibitem{newmad}
	   J.~Alwall, P.~Demin, S.~de Vissher, R.~Frederix, M.~Herquet,
	   F.~Maltoni, T.~Plehn, D.~Rainwater, T.~Stelzer,
	   JHEP {\bfseries 0709} (2007) 028.
   \bibitem{nvidia}
	   \texttt{http://www.nvidia.com/page/home.html}

   \bibitem{bases}
	   S.~Kawabata, Comput.~Phys.~Commun.
	   {\bfseries 41}(1986) 127. 

   \bibitem{pdflib}
	   H.~Plothow-Besch,
	   Comput.\ Phys.\ Commun. {\bfseries 75} (1993) 396,
	   Int.\ J.\ Mod.\ Phys.\ {\bfseries A10} (1995) 2901.
  \end{thebibliography}
  %

\end{document}